\documentclass[9pt,twocolumn,twoside]{opticajnl}
\journal{opticajournal} 

\setboolean{shortarticle}{true}

\usepackage{lineno}

\title{Temporal soliton generation in an ultra-high-effective-Q Kerr resonator enabled by Raman gain}

\author[1,*]{G.~Semaan}
\author[1]{Y.~Sun}
\author[1]{N.~Englebert}
\author[1]{S.~P.~Gorza}
\author[1]{F.~Leo}

\affil[1]{OPERA-photonics, Universit\'e libre de Bruxelles, 50 Ave. F. D. Roosevelt, B-1050 Brussels, Belgium}

\affil[*]{georges.semaan@ulb.be}

\begin{abstract}
We demonstrate temporal pattern formation in a coherently driven fiber ring cavity whose effective finesse is continuously reconfigured using distributed Raman amplification. We achieve an effective finesse of up to $\mathcal{F}_{\mathrm{eff}}\approx800$, corresponding to a linewidth of approximately 725 Hz ($Q\approx2.7\times10^{11}$) at 1555 nm. By exploiting the resulting increase in effective photon lifetime, we excite stable temporal cavity solitons and generate a low-repetition-rate frequency comb with a spacing of 580~kHz. Finally, we analyze the impact of the Raman loss-compensation mechanism, particularly its associated noise and show that a trade-off exists between soliton excitation threshold and stability.

\end{abstract}

\setboolean{displaycopyright}{false} 

\begin{document}

\maketitle



Temporal cavity solitons (CSs) are localized pulses sustained by a balance between dispersion and Kerr nonlinearity under coherent driving~\cite{Leo2010,Herr2014}. Over the past decade, these sources have attracted considerable interest and enabled a wide range of applications.

A major limitation, however, stems from the local energy exchange in resonators governed by the seminal Lugiato–Lefever (LL) equation \cite{Lugiato1987, haelterman_1992}. Because the Kerr response is nearly instantaneous and the dispersion is weak, the energy sustaining a cavity soliton must be supplied within its own temporal duration, resulting in high continuous-wave driving power thresholds. Synchronous pulsed driving can alleviate this constraint by reducing the average power injected into the cavity for a given peak power~\cite{Obrzud2017}. However, it increases experimental complexity, may introduce additional noise, and offers limited benefits for resonators operating at high repetition rates.
These fundamental limitations of LLE-class resonators have recently triggered extensive research efforts aimed at redistributing energy within the system~\cite{zhang_2024}. For instance, linear mode coupling has been exploited to reduce the oscillation threshold and achieve very high conversion efficiencies~\cite{Yang_eLight_2024, Helgason_NatPhoton_2021, helgason_2023, Hu_NatPhoton_2022}.

An alternative consists in providing the soliton with multiple energy sources. In this context, introducing intracavity gain enables the realization of active Kerr resonators, thereby relaxing some of the constraints imposed by passive cavities.This concept has been experimentally demonstrated in erbium-doped resonators~\cite{Englebert2021,Yao2024} and quantum cascade laser cavities~\cite{Columbo2021,Opacak2024,kazakov_2025}, yielding reduced soliton thresholds and enhanced conversion efficiencies at the cost of accumulated amplified spontaneous emission (ASE).

Raman gain is an attractive option as a gain mechanism for active cavities due to its broadband nature, lack of material doping requirements, and favorable saturation characteristics~\cite{Agrawal,Boyd}. While Raman amplification is a cornerstone of loss compensation in long-haul telecommunications \cite{Mollenauer1988, Islam2002} and enables direct soliton generation in various configurations~\cite{Li_nature_NZ,stokes_vahala}, its influence on the dynamics of coherently driven resonators has only recently begun to be explored. Raman gain in copropagating scheme has recently been used in silica microresonators to reduce effective loss, improve conversion efficiency, lower soliton thresholds, and to engineer transitions in Raman Kerr comb dynamics~\cite{Li2025,Jin2025}.

\begin{figure*}[htbp]
  \centering
  \includegraphics[width=\textwidth]{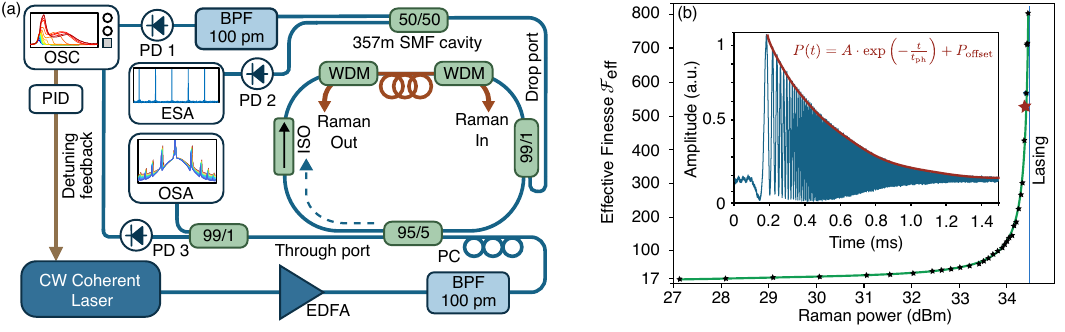}
  \caption{(a). Experimental setup. CW laser: 1555 nm laser, EDFA: Erbium doped fiber amplifier, BPF: Optical bandpass filter,
  PC: polarization controller, ISO: polarization independent dual stage isolator, 
  WDM: wavelength division multiplexer, Photodetectors PD1, PD3: 200 kHz , PD2: 1 GHz, PID: servo controller, OSA: optical spectrum analyzer 20 pm resolution, OSC: 1 GHz oscilloscope, ESA: electronic spectrum analyzer 43.5 GHz.
  (b) Linear cavity characterization: drop-port normalized transmission vs Raman pump power, showing the progressive increase of the effective finesse. The solid curve is a guide to the eye. The inset illustrates a single cavity ringdown measurement, corresponding to a single data point on the effective finesse curve. }
  \label{fig:setup}
\end{figure*}

In this work, we investigate Raman amplification as a loss compensation mechanism enabling nonlinear pattern formation, specifically temporal cavity solitons. 
Using a low-finesse cavity, for which soliton excitation is experimentally difficult to achieve under continuous-wave driving, we demonstrate that distributed Raman amplification can serve as a tunable control parameter for accessing the nonlinear regime. We define an effective finesse and connect it to soliton existence, spectral structure, and noise-limited stability.

The experimental setup is shown in Fig.~\ref{fig:setup}(a). The resonator consists of a 357~m single-mode fiber (SMF-28) ring, with a free spectral range (FSR) of 580.25~kHz and a total group-velocity dispersion of $\beta_2L=-7.85$~ ps$^2$ at 1555~nm. 
Raman amplification is implemented using a counter-propagating pump at 1455~nm with a spectral width of approximately 200~GHz, which is injected into and extracted from the ring via wavelength-division multiplexers (WDMs). In the absence of Raman pumping, the roundtrip power loss is $\approx36\%$, corresponding to a passive finesse of $\mathcal{F}\approx17$.

The cavity is coherently driven by a narrow-linewidth (Koheras Adjustik) continuous-wave (CW) laser at 1555~nm, which is amplified to 56~mW and coupled into the resonator through a 95/5 input coupler. A 99/1 output coupler provides access to the intracavity field for spectral and temporal characterization, while an optical isolator suppresses stimulated Brillouin scattering. The detuning between the CW laser and a cavity resonance is actively stabilized using a frequency-shifted control signal injected on the orthogonal polarization mode together with a servo controller. 
Furthermore, the cavity is placed in a temperature controlled box (Silentsys) to minimize thermally induced phase variations.
Temporal cavity solitons are individually excited via cross-phase modulation from an auxiliary single-pulse source~\cite{Leo2010}, which is not shown in Fig.~\ref{fig:setup}.

To quantify the impact of Raman amplification on the effective photon lifetime $t_{\text{ph}}$ and quality factor $Q_{\mathrm{eff}}$, we perform cavity ringdown measurements at low driving powers ($\approx600~\mu W)$~\cite{Dumeige2008,Huet2016}. Sweeping the laser frequency faster than the cavity decay rate excites a transient "ringing" response [see Fig.~\ref{fig:setup}(b) inset]. For each Raman pump power, we extract the intracavity energy decay by demodulating this signal via a Hilbert transform; fitting the resulting envelope with $e^{-t/t_{\text{ph}}}$ yields the effective photon lifetime. Unlike the cold cavity lifetime, this $t_{\text{ph}}$ accounts for the competition between intrinsic passive losses and active Raman gain. From $t_{\mathrm{ph}}$, we extract the effective roundtrip loss $\Lambda_{\mathrm{eff}} = t_R / t_{\mathrm{ph}}$ , where $t_R = 1 / \mathrm{FSR} \approx 1.72~\mu\text{s}$ is the cavity roundtrip time, and the corresponding effective finesse $\mathcal{F}_{\rm eff} = 2\pi / \Lambda_{\rm eff}$.
Figure~\ref{fig:setup}(b) shows the evolution of $\mathcal{F}_{\text{eff}}$ versus Raman pump power. While the passive cavity exhibits $\mathcal{F} \approx 17$, Raman amplification extends $t_{\text{ph}}$ by over an order of magnitude, reaching $\mathcal{F}_{\text{eff}} \approx 800$. This corresponds to an effective resonance linewidth $\Delta\nu=\mathrm{FSR} / \mathcal{F_{\rm eff}}\approx 725$\,Hz and an effective $Q_{\mathrm{eff}}=\nu_0 / \Delta\nu\approx 2.7\times 10^{11}$ at $\lambda_0=1555$~nm. 

\begin{figure}[t]
  \centering
  \includegraphics[width=\linewidth]{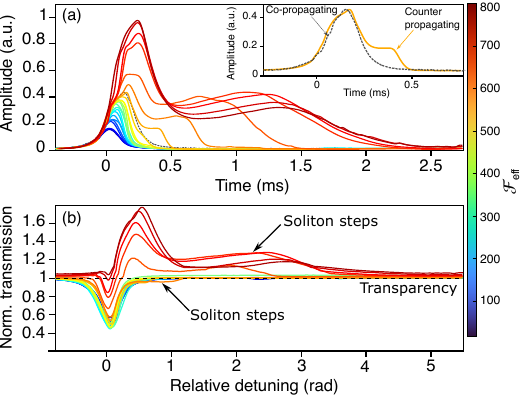}
\caption{Laser detuning scans at the (a) drop and (b) through ports show that increasing Raman pump power (at fixed driving power 56 mW) widens soliton steps, reflecting increased effective finesse.  Inset in (a): Dashed lines represent co-propagating resonances and solid lines counter propagating resonances at a pump power of 34.37~dBm.}
  \label{fig:detuning}
\end{figure}

We proceed to the nonlinear regime by increasing the driving power, with the aim of exploiting the resulting enhancement of the effective finesse to study pattern formation, particularly temporal cavity solitons.
The photon lifetime is a key parameter in Kerr nonlinear resonators, as it governs both the effective interaction length and the intracavity field build-up. Importantly, the Raman gain spectrum in silica ($\sim 5~\mathrm{THz}$, 40nm) is much broader than the spectral width of cavity solitons in long fiber resonators (typically a few nanometers). As a result, the Raman response can be treated as effectively instantaneous on the soliton timescale~\cite{Agrawal_NFO}.

Moreover, the counter-propagating geometry provides high saturation power such that, over the range of powers considered here (see below), the Raman gain is expected to be independent of the intracavity power.

\begin{figure*}[htpb]
  \centering
  \includegraphics[width=\textwidth]{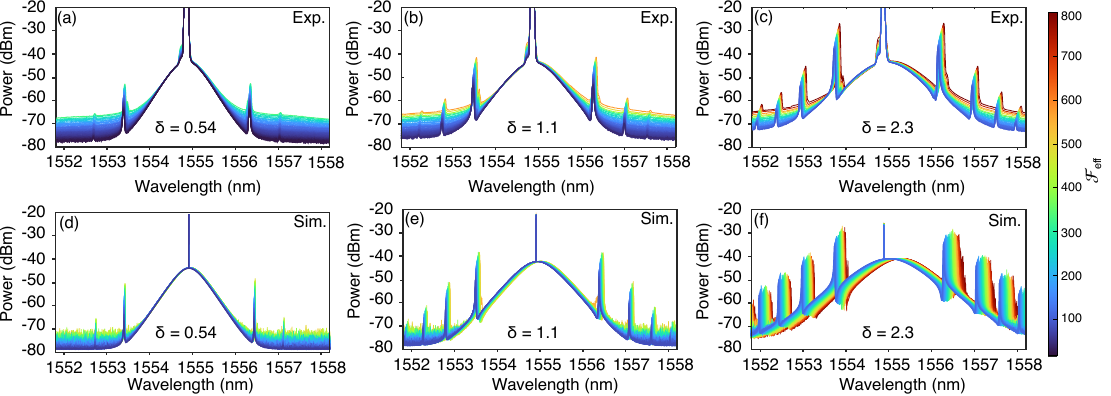}
\caption{Evolution of the temporal cavity soliton optical spectrum as a function of the effective finesse at different  detunings: $\delta = 0.54$~rad,  $\delta = 1.1$~rad and $\delta = 2.3$~rad.
(a-c) Experimental spectra measured at increasing Raman pump powers, corresponding to an effective finesse sweep. Stable soliton operation is observed over effective finesse ranges of 64–374 for $\delta = 0.54$ rad, 82–534 for $\delta = 1.1$ rad, and 131–790 for $\delta = 2.3$ rad.
(d-f) Corresponding numerical simulations based on the Ikeda-map model.
}
  \label{fig:specfinesse}
\end{figure*}

Fig.~\ref{fig:detuning} shows detuning scans at through and drop ports for increasing 1455~nm pump power with coherent driving power fixed at 56~mW.
The cavity is scanned at $\sim165~\mathrm{MHz/s}$ and the transmitted power is recorded with a 30~kHz detection bandwidth. At low Raman pump power, the resonance remains close to its unperturbed Lorentzian shape. As the effective finesse increases, Kerr-induced bistability tilts the resonances and broadens the detuning range supporting stable cavity solitons, manifesting experimentally as a widening soliton step~\cite{Coen2013Scaling,Herr2014}. At high Raman pump power, the drop-port resonance remains essentially similar to that of a passive resonator, whereas the through-port response evolves from a dip to a peak above the off-resonant baseline set by the driving field, indicating that the net round-trip gain approaches the cavity loss and that the cavity acts as an amplifier near resonance~\cite{Dumeige2008}.
As a point of comparison, we also show in Fig.~\ref{fig:detuning} the cavity transmission in the case of a Raman pump propagating in the opposite direction (co-propagating with the signal). We do not see any signature of soliton or pattern formation in that configuration likely because of a much stronger pump depletion.

We next analyze stationary patterns in our cavity, specifically, we investigate the impact of strong Raman pumping on the properties of temporal cavity solitons. 
To study the intrinsic properties of the soliton without interference from dynamics induced by the parameter scan, we excite a single cavity soliton at 56~mW and record its spectra while varying the Raman pump power at a fixed detuning. Figure~\ref{fig:specfinesse} shows the spectra obtained for $\delta = 0.54$, $\delta = 1.1$~rad, and $\delta = 2.3$, together with numerical simulations.
To model the intracavity field dynamics, we use a generalized Ikeda map. The evolution within each cavity round trip is governed by the generalized nonlinear Schr\"{o}dinger equation (GNLSE), accounting for chromatic dispersion ($\beta_2 = -22 \times 10^{-3}$ ps$^2$/m), Kerr nonlinearity ($\gamma = 1.3 \times 10^{-3}$ W$^{-1}$m$^{-1}$), and a delayed Raman response~\cite{Dudley}.
We stress that the Raman response is here centered at the soliton frequency and is different from the response used for loss compensation, which we approximate by a pump power dependent distributed gain along the fiber (see below). 
At each roundtrip, the intracavity field satisfies the boundary condition:
\begin{equation}
    A_{m+1}(T, 0) =i \sqrt{\theta_\text{in} P_d} + \exp\left(-i\delta - \Lambda_c/2\right) A_{m}(T, L),
\end{equation}
where $A$ represents the slowly varying field envelope, $\delta$ is the phase detuning, $\Lambda_c=2\pi/\mathcal{F}$ is the cold cavity loss and $\theta_\text{in} = 0.05$ is the power coupling coefficient for the pump $P_d$. 
To model finesse tuning while keeping the intrinsic cavity loss $\Lambda_c$ fixed, we introduce a distributed Raman gain with coefficient $g$ and treat it as broadband and unsaturated. This gain reduces the net effective round-trip loss to $\Lambda_{\rm eff}=\Lambda_c-gL$, so that the corresponding effective finesse is $\mathcal{F}_{\rm eff} = 2\pi / \Lambda_{\rm eff}$. 
Stochastic noise is injected at each round trip to account for quantum fluctuations and technical noise sources. 
To reproduce a qualitatively similar evolution of the experimental noise floor, we model the injected noise as a zero-mean complex Gaussian noise with variance
$
\sigma_q^2=\hbar\omega_0\,\eta_\text{exc},
$
where $\eta_\text{exc}=4$ is an excess-noise factor introduced as a parameter to emulate amplified spontaneous emission and other noise sources in a simplified manner~\cite{dudley_coen}.
We stress that this added noise is not intended as a fully physical model of the underlying processes; rather, it provides a controlled way to examine how the background noise level is modified by detuning and finesse.

Overall, we find good agreement between the experimental results and the simplified model. We only show simulation results over the range where we obtained stable solitons experimentally.
The low finesse threshold fits well with cavity soliton theory. The minimum finesse required to excite a soliton at a fixed detuning scales as $\mathcal{F}_{\mathrm{th}}\propto \sqrt{\delta}$ (see e.g. ref~\cite{englebert_high_2023} for the full expression). The origin of a high finesse limit below the lasing threshold (at low detuning) however, is left for future work and is likely due to experimental constraints. 
The solitons have durations of 1.7 to 4 ps and peak powers of 8 to 15 W. This corresponds to low average power and efficiency in the single soliton regime but the high effective finesse allows excitation with much lower coherent drive power. For example, the required driving power without loss compensation at $\delta=2.3$ is 3.6 W.

The most striking feature of the soliton spectra is the large amplitude of the Kelly-like sidebands (dispersive waves) emitted by the soliton, particularly at large detunings.
In our configuration, the losses are concentrated at the input and output couplers, while the Raman gain is distributed along the fiber, so that the intracavity power experiences a periodic modulation at the cavity round-trip frequency.
Such periodic perturbations are known to resonantly couple solitons to dispersive waves, generating sidebands at well-defined spectral offsets, in agreement with the soliton resonance picture described in Refs.~\cite{Hasegawa1990,Haseg1991}.
The increase in sideband amplitude with detuning is attributed to the higher spectral density at the Kelly sideband frequencies, while the increase with effective finesse is consistent with the longer effective photon lifetime.

We also observe a systematic dependence of the soliton self-frequency shift on the effective finesse. This behavior is consistent with cavity soliton theory~\cite{vahala_theory_2016, EngSimon} and is well reproduced by the simulations. 
In addition, the higher effective finesse is accompanied by a significant noise background associated with Raman amplified spontaneous emission. This noise increases sharply as the net round-trip gain approaches the laser threshold, following the inverse dependence predicted by standard rate equation models~\cite{Silfvast2008}.

\begin{figure}[t]
  \centering
  \includegraphics[width=\linewidth]{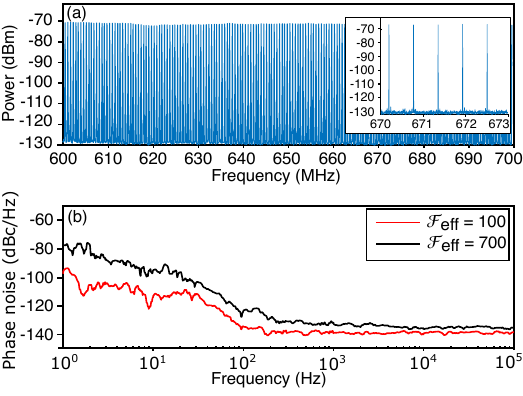}
  \caption{(a) Electrical spectrum of the soliton repetition rate within a bandwidth of 100 MHz at $\delta = 1.1$~rad and $\mathcal{F}_{\rm eff} = 300$.
(b) Single-sideband (SSB) phase noise of the soliton repetition rate, measured at 622~MHz (1073rd harmonic) and rescaled to the fundamental frequency, for two different effective finesse values at a fixed detuning $\delta = 1.1$~rad.}
  \label{fig:phasenoise}
\end{figure}

This constitutes the key difference from a passive resonator, since gain-induced noise directly impacts source performance. Although the behavior is well reproduced by the simulations, we stress that amplitude and phase fluctuations of the Raman pump are neglected and may introduce additional noise. To characterize the noise performance of our configuration, we measure the electrical spectrum of the single-soliton source. Figure~\ref{fig:phasenoise}(a) shows the electrical spectrum obtained for an effective finesse $\mathcal{F}_{\rm eff} = 300$.
This measurement does not indicate strong phase fluctuations originating either from amplified spontaneous emission or from Raman pump fluctuations, as the electrical spectrum exhibits more than 60~dB signal-to-noise ratio.
We also measured the phase noise of the $1073^{\mathrm{rd}}$ beat note (normalized to the fundamental) for two Raman pump powers at $\delta = 1.1$ [see Figure~\ref{fig:phasenoise}(b)], which indicates that the stability of the pulse train does depend on the Raman pump power, as expected due to the increased noise background.
A more comprehensive analysis of noise is beyond the scope of this Letter, but our results likely highlight a trade-off: increasing the effective finesse reduces the soliton excitation threshold but can degrade stability. Therefore, the preferred effective finesse depends on the available coherent driving power and the required noise performance.

In conclusion, we have shown that Raman gain can be used to transform a standard lossy fiber resonator into a reconfigurable ultra-high effective-$Q$ nonlinear cavity.
By tuning the counter-propagating Raman pump power, we continuously vary the effective finesse from $\mathcal{F}\approx 17$ to $\mathcal{F}_{\mathrm{eff}}\approx 800$ in the same 357~m SMF ring, reaching a photon lifetime of 22~ms ($Q_\mathrm{eff} \approx 3\times10^{11}$).
In this high-$Q$ regime, the cavity supports low-threshold temporal cavity solitons formation.
In future work, we will investigate more thoroughly the impact of Raman pumping on the soliton dynamics and on the range of existence of the soliton states. In particular, the role of pump coherence deserves further attention. 

Moreover, gain saturation and higher-order dispersion, neglected here, are expected to be important for broadband and high-efficiency comb generation and should also be addressed in future studies.

\begin{backmatter}
\bmsection{Funding}
MSCA (101150387, 101149506, 101103780), ERC (No.\ 101125625), and FWO and F.R.S.-FNRS (EOS, 40007560).


\bmsection{Disclosures}
The authors declare no conflicts of interest.

\bmsection{Data availability}
Data underlying the results presented in this paper may be obtained from the authors upon reasonable request.

\end{backmatter}

\bibliography{references}
 \bibliographyfullrefs{references}

\end{document}